\documentclass[twocolumn, tighten, times]{aastex63}

\usepackage{amsmath, amssymb, amsfonts}
\usepackage{graphicx}
\usepackage[dvipsnames, svgnames]{xcolor}
\usepackage{bm}
\usepackage{enumitem}
\usepackage{comment}
\usepackage{framed}
\usepackage{soul}
\usepackage{subfigure}

\begin{document}

\title{Turbulence Mode Decomposition and Anisotropy in Magnetically Dominated Collisionless Plasmas}

\author[0000-0002-3738-8980]{Samuel T. Sebastian}
\affiliation{Department of Physics, University of Florida, 2001 Museum Rd., Gainesville, FL 32611, USA}
\email{sebastian.s@ufl.edu}

\author[0000-0002-0458-7828]{Siyao Xu}
\affiliation{Department of Physics, University of Florida, 2001 Museum Rd., Gainesville, FL 32611, USA}
\email{xusiyao@ufl.edu}

\author[0000-0002-8455-0805]{Yue Hu}
\affiliation{Institute for Advanced Study, 1 Einstein Drive, Princeton, NJ 08540, USA}
\email{yuehu@ias.edu}

\author[0000-0001-8822-8031]{Luca Comisso}
\affiliation{Department of Physics, Columbia University, New York, NY 10027, USA}
\affiliation{Department of Astronomy, Columbia University, New York, NY 10027, USA}
\email{luca.comisso@columbia.edu}

\author[0000-0001-5796-225X]{Saikat Das}
\affiliation{Department of Physics, University of Florida, 2001 Museum Rd., Gainesville, FL 32611, USA}
\email{saikatdas@ufl.edu}

\author[0000-0002-3226-4575]{Joonas N\"attil\"a}
\affiliation{Department of Physics, University of Helsinki, P.O. Box 64, University of Helsinki, FI-00014, Finland}
\email{joonas.nattila@helsinki.fi}

\begin{abstract}
We use the 3D fully kinetic simulation to study different turbulence modes and turbulence anisotropy of relativistic turbulence in 
magnetically dominated collisionless plasmas. 
We extend the method developed
by
\citet{CL02}
for decomposing non-relativistic magnetohydrodynamic (MHD) turbulence 
into Alfv\'en, fast, and slow modes to the regime of 
collisionless plasmas. 
We find that 
Alfv\'en and slow modes are anisotropic, following the 
\citet{Goldreich1995} scaling,
while 
fast modes are isotropic. 
We observe 
a larger kinetic energy fraction of fast modes compared to that in the non-relativistic MHD turbulence, 
suggesting a stronger coupling of Alfv\'en and fast modes 
in relativistic magnetized turbulence in 
collisionless plasmas. 
We further examine the dynamic alignment 
and find a weaker scale dependence of the alignment angle than previously proposed.  
The dominant thermal fluctuations in the kinetic range 
can cause flattening of the turbulent velocity structure function 
and weakening of the turbulence anisotropy and 
dynamic alignment
near the kinetic scales. 

\end{abstract}


\section{Introduction} \label{sec:intro}
Most astrophysical plasmas are turbulent and magnetized \citep{Larson1981, Armstrong1995}. Turbulence plays an important role in energy transfer and dynamics in many processes in astrophysics and space physics, 
such as star formation \citep{McKee2007, Crutcher2012}, 
energetic particle transport in
the solar wind (e.g., \citealt{Chen2020, WHITMAN20235161}), 
and cosmic-ray transport and acceleration 
(e.g., \citealt{Fermi49,Yan_2008,Comisso_2024, Hu_2025}).

Significant progresses have been made in understanding nonrelativistic magnetohydrodynamic (MHD) turbulence in strongly collisional plasmas. 
The modern theory of strong MHD turbulence is established by \citet[][hereafter GS95]{Goldreich1995}.
The GS95 anisotropic scaling is only valid in the  
reference frame with respect to the local mean magnetic field 
\citep{Lazarian1999,Cho_2000},
rather than in the 
global reference frame with respect to the global mean magnetic field. 
It has been tested with extensive numerical studies with MHD simulations (e.g., \citealt{MaronGoldreich2001,Cho_2000,CLV02,Beresnyak_2015, Beresnyak_2014,2021ApJ...911...37H,2022MNRAS.512.2111H}). 

Decomposing MHD turbulent fluctuations into Alfvén, fast, and slow modes is important to understand the fundamental properties of compressible MHD turbulence and gain insight into the energy cascades and energy distributions of different modes 
\citep{CL02}. 
It is found that in nonrelativistic compressible MHD turbulence, Alfven modes exhibit the anisotropic scaling
similar to that in  incompressible MHD turbulence, and slow modes are passively mixed by the Alfvenic cascade \citep{LG01,CL02}. 
Fast modes evolve independently and have an isotropic energy distribution \citep{CL02}.
Extending the study with mode decomposition to the regime of relativistic turbulence in collisionless plasmas is thus essential for 
examining the validity 
of MHD turbulence theory in the new physical regime and the 
kinetic effects on mode interactions and energy transfer.

Plasmas in many high-energy astrophysical systems are strongly magnetized, weakly collisional, and have 
relativistic turbulent velocities.
Properties of relativistic magnetized turbulence have been previously studied with both MHD simulations (e.g., \citealt{Inoue_2011, Zrake_2012,Chernoglazov21,DelZanna2025}) and kinetic simulations (e.g., \citealt{Comisso_2018,Zhdankin2018a,Nättilä_2021, Sebastian_2025}). 
While the decomposition of magnetized turbulence into different modes has been explored in the regime of relativistic MHD turbulence \citep{Takamoto2016,Takamoto17}, it has not been applied to the relativistic magnetized turbulence in collisionless plasmas. 
In this work, 
we will use the fully kinetic 3D particle-in-cell (PIC) simulation, which provides a first-principle modeling of a collisionless plasma condition. 
We will perform turbulence mode decomposition in the regime of  relativistic turbulence in strongly magnetized collisionless plasmas. 
In particular, we will examine the turbulence anisotropy of different modes and the scale dependence of the anisotropy, in comparison with the GS95 theory.
We will also examine the alignment between the transverse turbulent velocity and magnetic fluctuations, in comparison with 
the theory of scale-dependent dynamic alignment \citep{Boldyrev2006}. 
In Section~\ref{sec:method}, we describe our numerical setup for the PIC simulation, as well as the MHD simulation as a comparison. 
In Section~\ref{sec:results}, we measure the velocity structure functions, perform turbulence mode decomposition, and analyze turbulence anisotropy. 
Finally, in Section~\ref{sec:conclusion}, we present our main conclusions. 

\begin{figure}
\begin{center}
   \includegraphics[width=8.65cm]{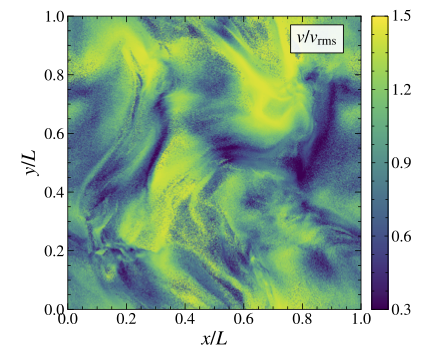}
   \includegraphics[width=8.65cm]{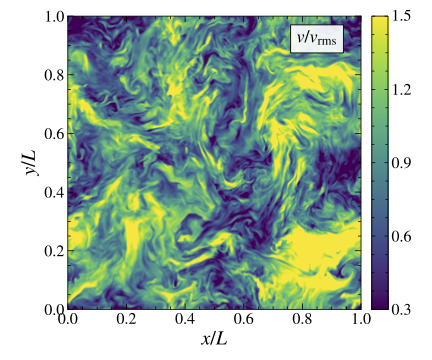}
\end{center}
\vspace{-0.5cm}
   \caption{Velocity field (normalized by the rms velocity $v_\text{rms}$)
   in the mid-z plane from the PIC (upper panel) and  MHD (lower panel) simulations. 
   }
\label{fig1}
\end{figure}

\section{Numerical Method} \label{sec:method}
\subsection{PIC Simulation}
\label{PIC_simulation}
{PIC simulations provide a first-principle approach for modeling magnetized turbulence in collisionless plasmas} 
\citep[e.g.,][]{Comisso_2018,Zhdankin2018a,Comisso_2019,Comisso_2021,Nättilä_2021,nattila2024}. 
The 3D PIC simulation of pair plasma analyzed in this study is performed  
with the \textsc{RUNKO} code \citep{Nattila_2022}. 
{The simulation is initialized with a uniform magnetic field $\bm{B_0}$
along the z axis.
The turbulent magnetic fluctuation is continuously driven such that the total magnetic field is $\bm{B}=B_0\mathbf{{ z}}+\delta \bm{B}$, where $\delta\bm{B}$ is the driven magnetic fluctuation in the direction perpendicular to the mean field, with the strength $\delta B\sim B_0$. 
The turbulence-driving scale is set equal to the numerical box size $L$ 
{to maximize the numerically-costly 3D inertial range}. 
We set the plasma magnetization parameter $\sigma_0=B_0^2/4\pi n_0 m_ec^2 =10$, where $n_0$ is the total number density of electrons and positrons, $m_e$ is the electron rest mass, and $c$ is the speed of light.
We consider a strongly magnetized plasma such that the Alfv\'en speed $v_A = c[\sigma_0/(1+\sigma_0)]^{1/2}$  
approaches $c$. 
The plasma is initialized with a dimensionless temperature of
$\Theta_0=k_BT_{0}/m_ec^2=0.3$, where $k_B$ is the Boltzmann constant, and $T_0$ is the initial plasma temperature. 
{The corresponding initial plasma $\beta_0$ (ratio of thermal pressure to magnetic pressure) is 
$\beta_0 = 2\Theta_0/\sigma_0 = 0.06$.}
The spatial resolution is $768^3$ cells, and the initial plasma skin depth $d_{e0}=c/\omega_{p,e}$ is resolved with 3 cells, where $\omega_{p,e}=\sqrt{4\pi n_0 e^2/m_e}$ is the plasma frequency, and $e$ is the electron charge. 
{We use 8 particles per cell per species to model the plasma.}
When the turbulence is fully developed, heating causes increase in plasma $\beta$ and 
the plasma skin depth $d_e$, 
with $\beta \approx 0.3$ 
{and $d_e \approx 5$ cells}
corresponding to the data analyzed in this work. }


\subsection{MHD Simulation}

The 3D isothermal MHD turbulence simulation analyzed in this study is generated using the AthenaK code \citep{2024arXiv240916053S,2024MNRAS.527.3945H}. It solves the ideal MHD equations with periodic boundary conditions, given by:
\begin{equation}
\label{eq.zeus}
\begin{aligned}
& \frac{\partial\rho}{\partial t} + \nabla \cdot (\rho \pmb{v})  = 0, \\
& \frac{\partial(\rho \pmb{v})}{\partial t} + \nabla \cdot \left[ \rho \pmb{v} \pmb{v}^T + (c_s^2\rho + \frac{B^2}{8\pi})I - \frac{\pmb{B}\pmb{B}^T}{4\pi} \right]  = \pmb{f},  \\
& \frac{\partial \pmb{B}}{\partial t} - \nabla \times (\pmb{v} \times \pmb{B}) = 0, \\
\end{aligned}
\end{equation}
where ${\bm v}$ is the fluid velocity, 
$c_s$ is the sound speed,
$\rho$ is the fluid density, ${\bm B}$ is the magnetic field, and 
$\pmb{f}$ is a forcing term. We also consider a zero-divergence condition $\nabla \cdot \pmb{B} = 0$, and an isothermal equation of state. 
The initial conditions include a uniform density $\rho_0$ 
and a uniform magnetic field ${\bm B_0}$
aligned along the z-axis. 
The turbulent kinetic energy is solenoidally and isotropically injected at the 
driving scale equal to the half numerical box size $L/2$. 
We continuously drive turbulence 
until 
a Kolmogorov energy spectrum is
fully developed. 
The turbulence sonic Mach number is $M_s = V_L/c_s\approx 1$, and the Alfv\'en Mach number is $M_A = V_L /V_A\approx 0.5$, where $V_L$ is the driven turbulent velocity at the driving scale, and $V_A = B_0/\sqrt{4\pi\rho_0}$ is the 
Alfv\'en speed. 
The corresponding plasma $\beta$ is $\approx 0.5$.
The simulation is gridded into $792^3$ cells, and  
the numerical dissipation scale is approximately $l_d \approx 10$ cells.


\begin{figure}[t!]
  \begin{center}
    \includegraphics[width=8.80cm]{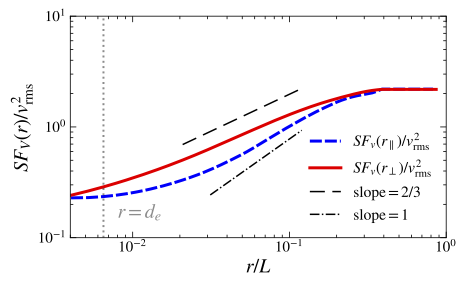}
    \includegraphics[width=8.80cm]{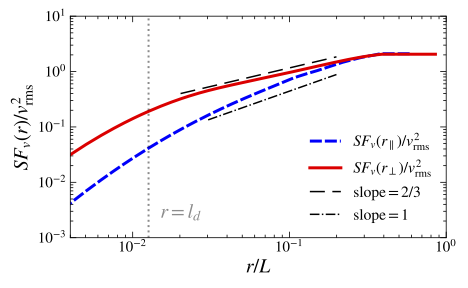}
  \end{center}
  \caption{$SF_v(r_\|)$ and $SF_v(r_\perp)$ (normalized by $v_\text{rms}^2$)
  measured in the PIC (upper panel) and MHD (lower panel) simulations. 
  The GS95 anisotropic scalings are indicated by the dashed and dash-dotted lines. The vertical dotted lines indicate $d_e$ in the PIC simulation and the numerical dissipation scale $l_d$ in the MHD simulation. 
  }
  \label{fig2}
\end{figure}


\begin{figure*}[t]
\centering
\includegraphics[width=8.65cm]{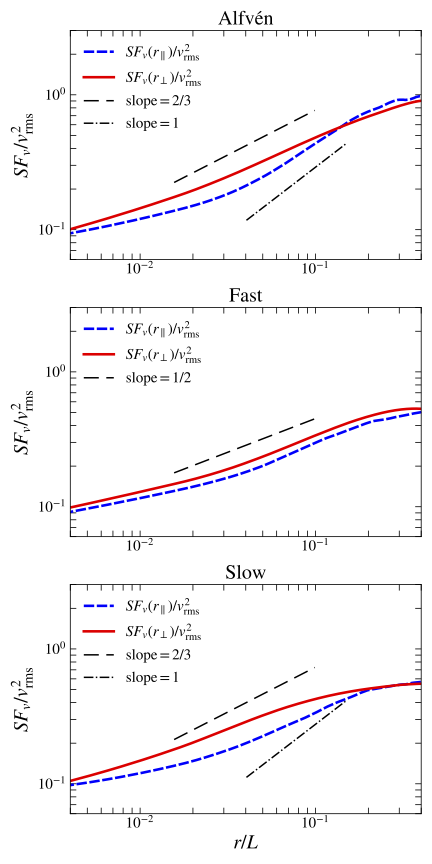}
\includegraphics[width=8.65cm]{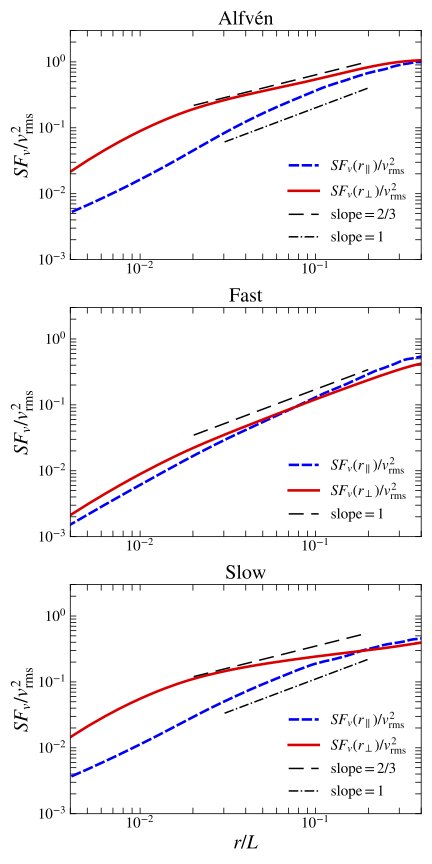}
\vspace{-0.5cm}
   \caption{$SF_v(r_\|)$ and $SF_v(r_\perp)$ of different turbulence modes for the PIC (left) and MHD (right) simulations. }
\label{fig3}
\end{figure*}


\begin{figure}
\begin{center}
   \includegraphics[width=8.65cm]{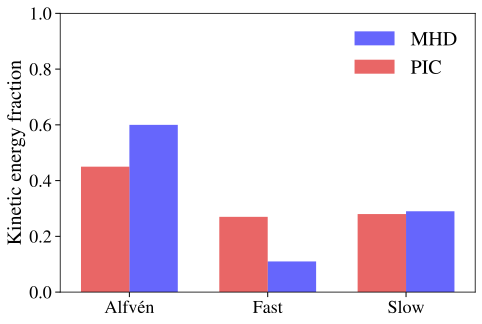}
\end{center}
\vspace{-0.5cm}
   \caption{Kinetic energy fractions of different turbulence modes measured from the PIC and MHD simulations. 
   }
\label{fig4}
\end{figure}

\begin{figure*}[t!]
  \centering
  \includegraphics[width=0.49\linewidth]{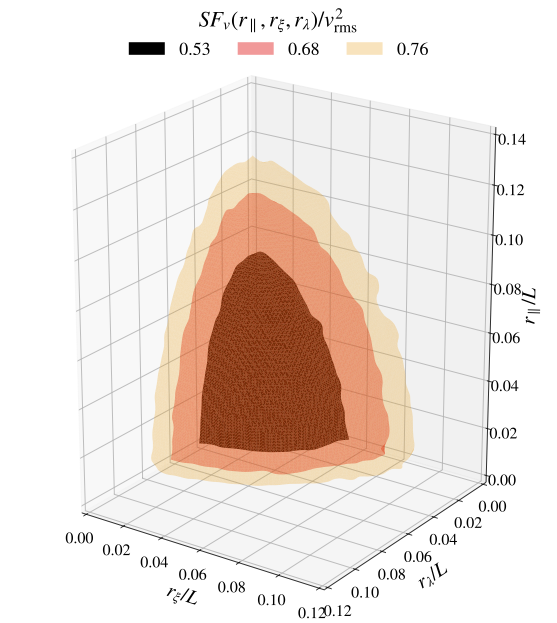}
    \includegraphics[width=0.49\linewidth]{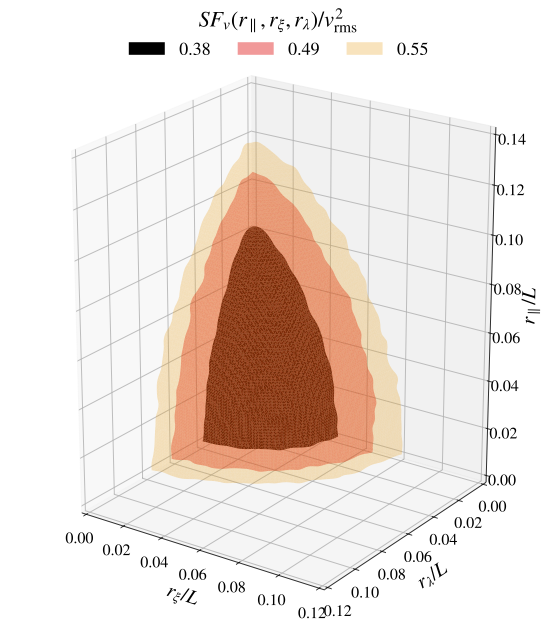}

  \caption{$SF_v(r_\|,r_\xi,r_\lambda)$ for the PIC (left panel) and MHD
  (right panel) simulations. 
  }
  \label{fig5}
\end{figure*}


\section{Results}\label{sec:results}

\subsection{
{Anisotropy of magnetized turbulence}} \label{sec:sf}

As a comparison, Fig. \ref{fig1} shows the velocity field from both simulations in the plane perpendicular to the mean magnetic field. 
In the PIC simulation, the velocity field is defined as the moment of the particle distribution function, obtained by averaging particle velocities. Despite the different driving methods, 
{both methods preferentially drive Alfv\'enic turbulence, and}
similar large-scale turbulent velocity fluctuations are developed in both simulations. 
Due to the heating in the PIC simulation, the thermal fluctuations become prominent at small scales and disrupt the small-scale turbulent velocity structures.
The similar effect of thermal fluctuations on hydrodynamic turbulence has been studied 
(e.g., \citet{eyink2021dissipationrangefluidturbulencethermal,Ma2023,Liu2026}).
In the isothermal MHD simulation, we see small-scale 
Alfv\'enic shear motions transverse to the magnetic field.


By following the method described in \citet{Cho_2000},
we measure the second-order structure function (SF) of velocity fluctuations with respect to the local mean magnetic field.
For each pair of points at positions $\boldsymbol{r}_1$ and $\boldsymbol{r}_2$, we define the local mean magnetic field as
\begin{equation}
\boldsymbol{B}_{\ell} = \tfrac{1}{2}\big[\boldsymbol{B}(\boldsymbol{r}_1)+\boldsymbol{B}(\boldsymbol{r}_2)\big], 
\qquad
\hat{\boldsymbol{z}} = \boldsymbol{B}_{\ell}/|\boldsymbol{B}_{\ell}|.
\end{equation} 
The SF of velocity fluctuations is defined as
\begin{align}
SF_v(r_\|,r_\perp) &= \big\langle \,(\boldsymbol{v}(\boldsymbol{r}_2)-\boldsymbol{v}(\boldsymbol{r}_1))^2 \,\big\rangle,
\end{align} 
where $r_\parallel=|\hat{\boldsymbol{z}} \cdot (\boldsymbol{r}_2-\boldsymbol{r}_1)|$ and $r_\perp=|\hat{\boldsymbol{z}} \times (\boldsymbol{r}_2-\boldsymbol{r}_1)|$ 
are the parallel and perpendicular components of the separation
$r=|\boldsymbol{r}_2-\boldsymbol{r}_1|$, respectively, 
and $\langle...\rangle$ denotes a spatial average.

In 
Figure~\ref{fig2},  
we present the measured $SF_v(r_\|,0)$ (hereafter $SF_v(r_\|)$)
and $SF_v(0,r_\perp)$ 
(hereafter $SF_v(r_\perp)$)
for the fully developed turbulence 
in the PIC and MHD simulations. 
The GS95 model with 
$SF_v(r_\perp)\propto r^{2/3}$ and $SF_v(r_\|)\propto r^1$ scalings are indicated as a reference. 
Despite {the transverse driving in the PIC simulation, with $\delta B\sim B_0$ at the driving scale,}
isotropy near the driving scale is also seen in the PIC simulation. 
Within the inertial range of turbulence with $SF_v(r_\perp)\propto r^{2/3}$ following the Kolmogorov slope, the GS95 scale-dependent anisotropy is seen in both cases. 
The turbulent energy cascade is mainly in the direction perpendicular to the magnetic field. 
Down to smaller scales, the steepening of $SF_v$ in the MHD simulation is caused by the numerical dissipation, while the flattening of $SF_v$ in the PIC simulation
near $d_e$
is due to the growing thermal fluctuations. 
The result from the MHD simulation agrees with earlier studies on compressible sub-Alfv\'enic MHD turbulence in low-$\beta$ plasmas 
\citep{CL02,Cho03}.
For the PIC simulation, 
we observe a weaker anisotropy compared to that in the MHD simulation. 
This is probably due to the 
larger $M_A \approx \delta B/B_0$
\citep{Lazarian1999}
and the 
significant heating effect at small scales in the PIC simulation. 
The isotropic thermal fluctuations reduce the turbulence anisotropy seen in the PIC simulation. 


\subsection{Anisotropy of Alfv\'en, fast, and slow modes} \label{sec:mode}
To further examine the anisotropy of different modes of magnetized turbulence, 
we decompose the turbulent velocity fluctuations into Alfv\'en, fast, and slow modes by following the method established in \citet{CL02}. 

In Fig.~\ref{fig3}, we present the measured $SF_v(r_\|)$ and $SF_v(r_\perp)$ for  different MHD modes in both PIC and MHD simulations. 
In both cases, Alfv\'en and slow modes exhibit anisotropy within the inertial range of turbulence, while fast modes are basically isotropic at all scales. 
The result from the MHD simulation is consistent with that in e.g., \cite{Cho03}.
Alfv\'en and slow modes in the MHD simulation follow the GS95 scalings. 
This is expected as Alfv\'en modes dominate the dynamics of MHD turbulence, and slow modes are passively mixed by Alfv\'en modes 
\citep{LG01}.
In non-relativistic MHD turbulence, 
fast modes have their independent energy cascade 
\citep{CL02}.
We see that the $SF_v$ of fast modes has a slope close to 1. This is consistent with the scaling of fast modes studied with compressible MHD simulations in e.g., \citet{Kowal_2010}.

In the PIC simulation, 
$SF_v$ of fast modes
is shallower than that in the MHD simulation, with a slope close to $1/2$, similar to that for the acoustic turbulence
\citep{CL02}. 
We also note that the GS95 scalings are better observed for Alfv\'en modes, while slow modes seem to be more subject to the heating effect, and their $SF_v$ is more flattened compared to Alfv\'en modes.


We also measure the kinetic energy fractions of different turbulence modes (see Fig. \ref{fig4}) and find $f_A \approx 0.45$, $f_f \approx 0.27$, and $f_s \approx 0.28$ for the energy fractions of Alfv\'en, fast, and slow modes 
measured in the PIC simulation, and $f_A \approx 0.60$, $f_f \approx 0.11$, and $f_s \approx 0.29$ for the MHD simulation. 
With the solenoidal driving adopted in the MHD simulation, the small $f_f$ indicates inefficient 
generation of fast modes due to the weak coupling between Alfv\'en and fast modes \citep{CL02}. 
The higher $f_f$ seen in the PIC simulation indicates a stronger coupling between Alfv\'en and fast modes in the relativistic magnetized turbulence. 
Previously, the strong coupling 
of Alfv\'en and fast modes in 
compressible relativistic turbulence has been found with MHD simulations
in magnetically dominated plasmas
\citep{Takamoto17}.


\subsection{Examining the dynamic alignment}
\label{sec:SF3D}

It is proposed 
in \citet{Boldyrev2006}
that the polarizations of Alfv\'enic velocity and magnetic fluctuations 
in the plane perpendicular to the magnetic field 
can be spontaneously aligned, and this alignment becomes stronger at smaller scales in a turbulent cascade. 
This polarization alignment would also  
lead to the anisotropy of turbulent velocities in the plane perpendicular to the magnetic field, and this anisotropy 
is expected to increase with decreasing scales.

To further examine the turbulence anisotropy in the plane perpendicular to the local mean magnetic field, 
we generalize $SF_v$ to distinguish between the two perpendicular directions, as 
$SF_v(r_\|,r_{\xi},r_{\lambda})$.
$r_\|$ remains a component of the separation $r$ parallel to the local mean magnetic field.
$r_\xi$ is one perpendicular component aligned with the transverse magnetic fluctuation 
\begin{equation}
   \delta \boldsymbol{b}_{\perp r} = \delta \boldsymbol{b}_r - ( \delta\boldsymbol{b}_r \cdot \hat{\boldsymbol{z}})\hat{\boldsymbol{z}}, ~~ \delta \boldsymbol{b}_r = 
    \boldsymbol{B}(\boldsymbol{r_2})-
    \boldsymbol{B}(\boldsymbol{r_1}),
\end{equation}
in the direction
$\hat{\boldsymbol{e}}_{\xi} = \delta \boldsymbol{b}_{\perp r}/|\delta\boldsymbol{b}_{\perp r}|$.
The other perpendicular component $r_\lambda$ is in the direction  $\hat{\boldsymbol{e}}_{\lambda} = \hat{\boldsymbol{z}} \times \hat{\boldsymbol{e}}_{\xi}$. 

Figure~\ref{fig5} shows the 
contours of equal $SF_v$ measured in the above local coordinate systems, representing the shapes of turbulent eddies. 
We see that smaller eddies are more elongated along the local magnetic field, showing the scale-dependent anisotropy predicted by the GS95. 
In the plane perpendicular to the local mean magnetic field, the anisotropy is much weaker, with a slight difference between $r_\xi$ and $r_\lambda$ for the same $SF_v$ contour. 

\begin{figure}
\begin{center}
   \includegraphics[width=8.65cm]{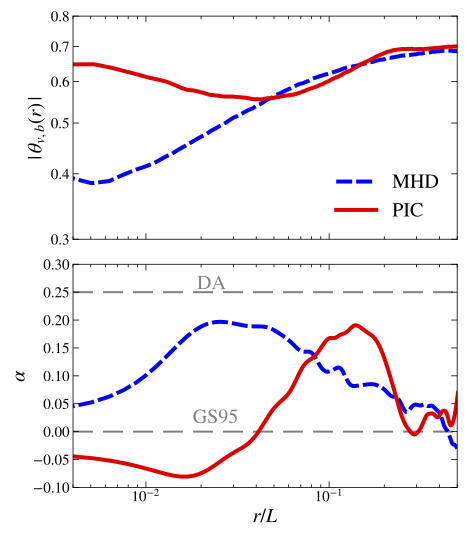}
\end{center}
\vspace{-0.5cm}
   \caption{Alignment angle $\theta_{v,b}(r)$ (upper panel) and alignment slope $\alpha$ (lower panel) measured in the PIC and MHD simulations. ``DA" represents the theoretical expectation in \citet{Boldyrev2006}. 
   }
\label{fig6}
\end{figure}

To further examine the polarization alignment and its scale dependence, we follow 
a similar method to that adopted in e.g.,
\citet{Mason_2006,Dong_2022,beattie2025}, 
and we use the reference frame with respect to the local mean magnetic field. The alignment angle is measured as 
\begin{equation}
     \theta_{v,b}(r) \sim \mathrm{sin}(\theta_{v,b}(r)) = \frac{\langle |\delta \boldsymbol{v}_{\perp r} \times \delta \boldsymbol{b}_{\perp r}| \rangle}{\langle |\delta \boldsymbol{v}_{\perp r}||\delta \boldsymbol{b}_{\perp r}|\rangle}.
\end{equation}
%
Here $\delta \boldsymbol{v}_{\perp r}$ is the transverse velocity fluctuation, 
\begin{equation}
  \delta \boldsymbol{v}_{\perp r} = \delta \boldsymbol{v}_r - ( \delta\boldsymbol{v}_r \cdot \hat{\boldsymbol{z}})\hat{\boldsymbol{z}}, ~~ \delta \boldsymbol{v}_r = \boldsymbol{v}(\boldsymbol{r}_2)-\boldsymbol{v}(\boldsymbol{r}_1).
\end{equation}
Figure~\ref{fig6} shows $\theta_{v,b}(r)$
and 
the slope $\alpha$ from the measured $\theta_{v,b}(r) \propto r^{\alpha}$ as a function of $r$ for the PIC and MHD simulations. 
In both simulations, we see a large $\theta_{v,b}(r)$, indicative of a weak alignment. 
This is consistent with the weak turbulence anisotropy in the plane perpendicular to the local mean magnetic field seen in Figure~\ref{fig5}.
There is nearly no polarization alignment near the driving scale with $\theta_{v,b}(r)\approx 0.7$. 
In the MHD simulation, $\theta_{v,b}(r)$
decreases with decreasing scales in both the inertial range and the numerical dissipation range, with a stronger alignment seen deeper in the numerical dissipation range. 
In the PIC simulation, $\theta_{v,b}(r)$ slightly decreases in the initial range, but increases toward the kinetic range, 
until the alignment is completely lost in the kinetic range
with 
$\theta_{v,b}(r)\approx 2/\pi$.
$\alpha=0.25$ corresponds to the theoretical expectation in \citet{Boldyrev2006} for Alfv\'enic turbulence. 
In both simulations, 
we find a weaker scale dependence of $\theta_{v,b}(r)$. 
$\alpha$ ranges from $\approx 0$ to $\approx 0.2$ and saturates near the numerical dissipation scale in the MHD simulation. 
In the PIC simulation, within the turbulence inertial range, $\alpha$ increases from $\approx 0$ to $\approx 0.2$ and then drops back to $\approx 0$. 
Toward the kinetic range, 
it further decreases and becomes negative. 
{The difference between the PIC and MHD simulations may be caused by the different driving methods, driving and plasma conditions, and dissipation mechanisms. 
Although we do not expect that the driving affects the turbulence anisotropy in the inertial range,
the driving method adopted in the PIC simulation can minimize the
transition from the driving scale to the inertial regime of the turbulence
\citep{TenBarge2014}}, and thus  
may cause $\alpha$ to reach its peak at a larger scale compared to the MHD simulation.  
The larger $\theta_{v,b}(r)$ and the negative $\alpha$ toward smaller scales 
in the PIC simulation is likely caused by the dominant thermal fluctuations in the dissipation range.


A stronger scale dependence
with $\alpha$ reaching $\approx 0.25$ is reported in 
\citet{Mason_2006}
with an incompressible MHD simulation of resolution $256^3$.
Their measurement is performed in the global reference frame with respect to ${\bf B_0}$, and $r$ is the separation in the plane perpendicular to ${\bf B_0}$.
We adopt the local reference frame with respect to the local mean magnetic field, as only the local magnetic field is relevant to the turbulent eddy at the scale of interest. 

Higher-resolution MHD simulations by \citet{Beresnyak_2012,Beresnyak_2015}
suggest that the alignment slope $\alpha$ decreases with increasing numerical resolutions. 
It converges to $\approx 0.06$ based on MHD simulations of Alfv\'enic turbulence
with resolutions up to $4096^3$
\citep{Beresnyak_2015}.
In the regime of relativistic turbulence in collisionless plasmas, 
for a systematic comparison with the theory of dynamic alignment, 
higher-resolution PIC simulations of Alfv\'enic turbulence 
will be needed for the convergence study. 
{In Appendix, we present the results from a lower-resolution PIC simulation, and a similar trend as in Fig. \ref{fig6} is seen.}


\section{Conclusions} \label{sec:conclusion}

We perform turbulence mode decomposition and investigate the
turbulence anisotropy of relativistic turbulence in 
strongly magnetized
collisionless pair plasma with the 3D PIC simulation. 
We focus on the dynamics of turbulence and thus the turbulent velocity fluctuations. 

We find the SF of total velocity fluctuations are consistent with the GS95 anisotropic scaling in the inertial range of turbulence, as also shown in 
\citet{Sebastian_2025}.
Previously, the SF of magnetic fluctuations consistent with the GS95 anisotropic scaling in relativistic turbulence in collisionless pair plasma modeled with 3D PIC simulations
is reported in 
\citet{Zhdankin2018a}.

In comparison with the isothermal MHD turbulence simulation with numerical dissipation, 
we find that thermal fluctuations are dominant in the kinetic range in the PIC simulation with the heating effect. 
They cause the flattening of $SF_v$ near the kinetic scales 
and reduce the turbulence anisotropy seen in the PIC simulation. 
By contrast, the numerical dissipation in the MHD simulation causes the steepening of 
$SF_v$ near the dissipation scale.

We further extend the method for MHD turbulence mode decomposition developed in 
\citet{CL02}
to relativistic turbulence in collisionless plasmas. 
Among the decomposed Alfv\'en, fast, and slow modes, 
we find that Alfv\'en and slow modes are anisotropic, while fast modes are isotropic. 
This result is similar to that found in nonrelativistic MHD turbulence simulations  
\citep{CL02,Cho03}. 
However, 
in comparison with the MHD simulation, the $SF_v$ of fast modes in the PIC simulation has a shallower slope, and the kinetic energy fraction of fast modes is higher. 
This indicates a stronger coupling between Alfv\'en and fast modes in relativistic  turbulence in strongly magnetized
collisionless plasmas. 
This finding is consistent with that in 
\citet{Takamoto17}
for relativistic MHD turbulence, 
but further studies are needed to exclude the driving effect
and examine the dependence of their coupling on the plasma magnetization. 

In both compressible PIC and MHD simulations, 
we find the polarization alignment between the turbulent velocity and magnetic fluctuations in the plane perpendicular to the local mean magnetic field is weak, and the maximum alignment slope $\alpha$ is less than the theoretical prediction in 
\citet{Boldyrev2006}.
In the MHD simulation, the alignment angle decreases within the inertial range and further decreases in the dissipation range. The alignment slope 
saturates at $\alpha \approx 0.2$ near the numerical dissipation scale, given the numerical resolution in this work. 
In the PIC simulation, the alignment angle slightly decreases in the inertial range but increases toward the dissipation range. 
The resulting negative $\alpha$ 
at smaller scales
is probably due to the heating effect in the PIC simulation.  
For a systematic test of the dynamic alignment theory
in the regime of relativistic turbulence in collisionless plasmas, 
higher-resolution 3D PIC simulations of Alfv\'enic turbulence 
with a more extended inertial range 
will be needed.

\section*{Acknowledgments}
S.S acknowledges support from the University of Florida’s
University Scholars Program. S.X. and S.D. acknowledge the support from the NASA
ATP award 80NSSC24K0896. 
L.C. acknowledges support from NSF award PHY-2308944 and NASA ATP grant 80NSSC24K1230.
J.N. acknowledges funding by the Research Council of Finland grant No.~372879.
S.S., S.X., and S.D. acknowledge
UFIT Research Computing for providing computational
resources and support contributing to the research results
reported in this work.

\bibliographystyle{aasjournal}
\bibliography{references}

\appendix

\begin{figure*}
    \begin{center}
   \includegraphics[width=17cm]{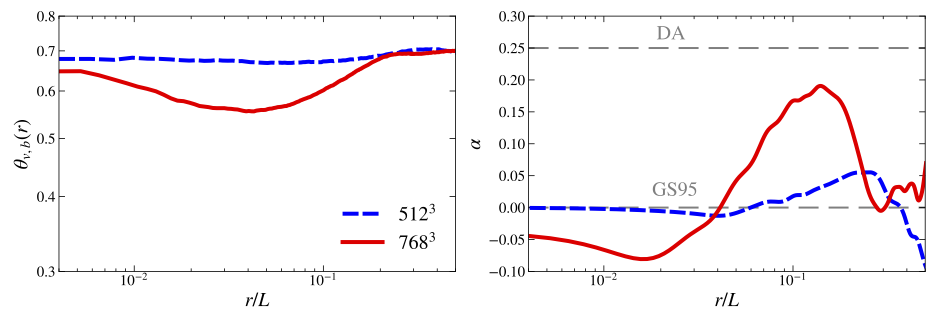}
    \end{center}
    \vspace{-0.5cm}
   \caption{Alignment angle $\theta_{v,b}(r)$ (left panel) and alignment slope $\alpha$ (right panel) measured in PIC simulations with different spatial resolutions. 
   }
\label{fig_appendix}
\end{figure*}

{We perform a lower-resolution $512^3$ 
PIC simulation with the same initial conditions as detailed in Section~\ref{sec:method}. 
We analyze the simulation when turbulence is fully developed, corresponding to $\beta \approx 0.6$, and $d_e \approx 5$ cells. 
Figure~\ref{fig_appendix} compares the alignment angle and slope measured from the higher-resolution PIC simulation as shown in Fig. \ref{fig6}
and that measured from the lower-resolution simulation. 
A similar trend is seen. 
Given the shorter turbulence inertial range at a lower resolution, $\theta_{v,b}(r)$ and $\alpha$ both vary within a smaller range and more weakly depend on $r$.}

\end{document}